\documentclass[conference]{IEEEtran}
\IEEEoverridecommandlockouts
\usepackage{cite}
\usepackage{amsmath,amssymb,amsfonts}
\usepackage{algorithmic}
\usepackage{graphicx}
\usepackage{textcomp}
\usepackage{xcolor}
\usepackage{comment}
\usepackage{color}
\usepackage{subcaption}
\usepackage{rotating} 
\usepackage{multirow}
\usepackage{color}
\usepackage{xcolor, colortbl}
\usepackage{lscape}
\usepackage{array, multirow, multicol}
\usepackage{amsmath} 
\usepackage{hyperref}

\definecolor{verd}{rgb}{0.5, 1.0, 0.0}

\newif\ifremark
\long\def\remark#1{
\ifremark%
        \begingroup%
        \dimen0=\columnwidth
        \advance\dimen0 by -1in%
        \setbox0=\hbox{\parbox[b]{\dimen0}{\protect\em #1}}
        \dimen1=\ht0\advance\dimen1 by 2pt%
        \dimen2=\dp0\advance\dimen2 by 2pt%
        \vskip 0.25pt%
        \hbox to \columnwidth{%
                \vrule height\dimen1 width 3pt depth\dimen2%
                \hss\copy0\hss%
                \vrule height\dimen1 width 3pt depth\dimen2%
        }%
        \endgroup%
\fi}    
    
\remarktrue

\def\BibTeX{{\rm B\kern-.05em{\sc i\kern-.025em b}\kern-.08em
    T\kern-.1667em\lower.7ex\hbox{E}\kern-.125emX}}
\begin{document}

\title{SYNPA: SMT Performance Analysis and Allocation of Threads to Cores in ARM Processors}

\author{


\IEEEauthorblockN{
Marta Navarro, 
Josué Feliu, 
Salvador Petit, 
María E. Gómez and 
Julio Sahuquillo
}

\IEEEauthorblockA{ 
Universitat Politècnica de València, Spain\\ 
\{marnaed,jofepre,spetit,megomez,jsahuqui\}@disca.upv.es
}

}


\maketitle

\begin{abstract}
Simultaneous multithreading processors improve throughput over single-threaded processors thanks to sharing internal core resources among instructions from distinct threads. However, resource sharing introduces inter-thread interference within the core, which has a negative impact on individual application performance and can significantly increase the turnaround time of multi-program workloads. The severity of the interference effects depends on the competing co-runners sharing the core. Thus, it can be mitigated by applying a thread-to-core allocation policy that smartly selects applications to be run in the same core to minimize their interference.

This paper presents SYNPA, a simple approach that dynamically allocates threads to cores in an SMT processor based on their run-time dynamic behavior. The approach uses a regression model to select synergistic pairs to mitigate intra-core interference.
The main novelty of SYNPA is that it uses just three variables collected from the performance counters available in current ARM processors at the dispatch stage. Experimental results show that SYNPA outperforms the default Linux scheduler by around 36\%, on average, in terms of turnaround time in 8-application workloads combining frontend bound and backend bound benchmarks.
\end{abstract}

\begin{IEEEkeywords}
Simultaneous Multithreading, Scheduling, High-Performance Computing, HPC applications, Thread-to-Core allocation, synergistic applications, intra-core interference
\end{IEEEkeywords}

\section{Introduction} 

Simultaneous multithreading (SMT)~\cite{TUL95} is the common multithreading paradigm implemented in recent high-performance server processors. This paradigm allows instructions from multiple threads to be issued in the same cycle, which helps increase the processor throughput. To make the design efficient, the major core resources (e.g., first-level caches and arithmetic operators) are shared among the co-running threads, increasing their utilization. In this way, throughput is improved with minimal area overhead. For instance, Marvell reports an area penalty of only 5\% to provide SMT support on the ThunderX3 processor while the performance gain 
is well over that value~\cite{ThunderX3}.

Resource sharing among applications executing in an SMT core causes inter-application interference that impacts the performance of individual applications. As the availability of core resources is key for performance, intra-core resource sharing can increase the execution time of individual applications significantly due to inter-application interference. Performance degradation, however, depends on the resource consumption of the applications co-running on the SMT core. Applications can present a wide diversity of resource demands. For instance, memory-bound applications present a high access rate to the memory hierarchy, while compute-intensive applications stress the 
arithmetic units. This means that the inter-application interference at the shared resources will strongly depend on the resource demands of the co-running applications.

As the performance of an application depends on the co-runner/s, a plausible approach to boost the processor performance is to rely on an \emph{intra-core interference-aware} thread allocation policy. 
This kind of policy selects the applications to be run on each core 
with the aim of minimizing the intra-core interference among them. 
This goal can be achieved, for instance, by selecting applications whose behaviors complement each other from a resource consumption perspective, e.g., a memory-bound application with a compute-intensive application. This work uses the term \emph{synergistic applications} to define applications that exhibit complementary behaviors.

Designing this kind of policy for real processors needs to tackle two main challenges.
The first is to determine if it is possible to characterize the \emph{synergy of applications} 
with the performance counters available in the target processor.
This is challenging as nowadays processors implement hundreds of performance counters working 
at different stages of the pipeline and measuring distinct magnitudes (e.g., cycles or events).
This challenge comprises two major decisions: at which processor stage should performance be analyzed and which performance counters from the ones available in the target processor must be used.
The second challenge is how the most synergistic pairs of applications are estimated. 

A few existing approaches have been proposed for Intel~\cite{top-down} and IBM~\cite{symbiotic-josue, symbiotic_tpds} processors. However, they cannot be applied to ARM processors because the performance counter architecture widely differs across them. Despite ARM processors gaining popularity in the data center segment thanks to their high energy efficiency \cite{the-adviser}, to the best of our knowledge, no existing thread-to-core approach can be applied to increase their throughput.

In this paper, we propose SYNPA, a thread allocation policy that selects synergistic applications and allocates them to the SMT cores of an ARM processor. SYNPA addresses the aforementioned first challenge by characterizing 
the performance of applications with a simple three-variable model at the dispatch stage. The second challenge is addressed by devising a linear regression model that predicts the performance of an application when running with another one in an SMT core.

Experimental results show that SYNPA outperforms the default Linux scheduler in terms of turnaround time by around 36\%, on average, in 8-application workloads combining frontend bound with backend bound applications and over 55\% in some workloads. SYNPA also improves the Linux scheduler in throughput and fairness by around 2.2\% and 25\%, respectively, across the same workloads.

This paper makes the following main contributions: 

\begin{itemize}
\item We present a simple model to estimate the performance of applications in SMT execution based on just three performance indicators from the dispatch stage.
\item The devised model presents extreme simplicity (only four performance counters), far from existing approaches developed for processors from other vendors ~\cite{symbiotic-josue}. This translates to a $40\%$ reduction in the overhead to estimate the performance of the possible combinations. 
\item To the best of our knowledge, this is the first \emph{feasible} thread allocation approach for an ARM processor with SMT cores.
\end{itemize}

\section{Related work}
One of the well-known weaknesses of SMT is that single-thread performance can be widely affected depending on the synergy of the applications executed simultaneously. 

Significant research works developed heuristics to address these weaknesses. Some approaches~\cite{Snavely_symbiotic,Acosta_2009} focus on generic processors using simulators. Snavely and Tullsen~\cite{Snavely_symbiotic} try to address this problem by periodically running a subset of the possible combinations to sample their performance. Cazorla et al.~\cite{Acosta_2009} show that system throughput relies on the scheduling policy regardless of the instruction fetch policy. Other approaches ~\cite{settle2004architectural,l1-aware, vega13,Radojkovic_2012, cal-marta} concentrate on commercial processors. ~\cite{settle2004architectural} uses offline profiling to improve processor throughput in an Intel Pentium-4 Xeon slightly. Feliu et al.~\cite{l1-aware} take scheduling decisions to balance the L1 bandwidth utilization among the cores in an Intel Xeon E5645. Also, using performance counters at runtime but focusing on multithread workloads, Vega et al.~\cite{vega13} present a heuristic to determine when threads should be consolidated in the SMT cores of the IBM POWER7 processor. Radojkovic et al.~\cite{Radojkovic_2012} propose a statistical inference method to estimate the task assignment in an UltraSPARC T2 processor. Navarro et al. \cite{cal-marta} propose a heuristic, based on the \cite{top-down} method, to identify \textit{symbiotic} pairs.
 
Other works establish linear regression models that use performance counters to estimate the performance of each combination of applications. Two early attempts ~\cite{Moseley, eyerman10} focused on a simulated generic and simplified processor. Moseley et al.~\cite{Moseley} employ linear modeling and recursive partitioning to estimate the speedup when executing two applications simultaneously. However, no specific performance counter is mentioned. Eyerman and Eeckhout~\cite{eyerman10} propose an analytical model that predicts the slowdown suffered by each application when co-scheduled with other applications. The proposed model is based on custom CPI stacks at the dispatch stage~\cite{Eyerman_CPIacc,Eyerman_SMTacc}; unfortunately, this approach is not practical as performance counters in existing processors can not collect the required events.

Finally, we discuss other works closer to ours using linear regression models built with performance counters available in real processors. In SMiTE~\cite{smite}, authors propose a regression model that combines \emph{sensitivity} and \emph{contentiousness} to estimate performance interference in an Intel Sandy Bridge processor. This work uses a single equation per core and 24 performance counters (12 for application) related to memory events like TLB and caches. Unfortunately, the prediction model performs poorly as the performance monitoring unit is only able to measure events for the entire core. Thus, there is no information for each individual thread running on the core.
Feliu et al.~\cite{symbiotic-josue} leverage the CPI accounting mechanisms of the IBM POWER8 and adapt the interference model to schedule the optimal combination of applications in the SMT cores. These mechanisms, however, are IBM specific and cannot be applied to processors from other vendors.

In addition to being the only approach that can work on performance counters of ARM processors, SYNPA is also simpler than the previous approaches proposed for Intel~\cite{top-down} and IBM~\cite{symbiotic-josue, symbiotic_tpds}. 
SYNPA requires only three equations and four performance counters to estimate how synergistic is the SMT execution of an application with a particular co-runner. Conversely, ~\cite{symbiotic-josue, symbiotic_tpds} propose a model with five equations (with the same number of floating-point multiplications and similar complexity) and six performance counters. Given that the number of possible combinations quickly explodes with the number of cores and applications, the reduction in the number of equations translates into a $40\%$ lower time (overhead) required to estimate the performance of all possible pairs of applications. 
Similarly, ~\cite{top-down} requires nine metrics that involve 15 performance counters to characterize the performance of an application.
In summary, little work has concentrated on regression models addressing the constraints imposed by real processors, where identifying the performance counters from the several hundred available in each specific platform to face the problem, if possible, imposes a real challenge. Moreover, despite the fact that ARM processors have recently emerged in the server market, to the best of our knowledge, this is the first work focusing on these processors.

\section{Characterizing Performance in ARM Processors}

This section discusses first general issues about measuring performance at a processor pipeline stage to devise a performance model. After that, we introduce the way we found to characterize the performance of ARM processors.

\subsection{General Issues and Architectural Constraints}

Instructions flow from outside the processor and enter into the pipeline at the fetch stage, traverse different stages, and leave the pipeline when they retire at the commit stage. Inside the pipeline, instructions flow in input (program) order at the starting stages (i.e., fetch, decode, and dispatch). After that, instructions are issued to the functional units for execution. At this point, the pipeline splits into multiple pipelines (e.g., integer ALU, floating-point unit, or load unit). Instructions are issued out of program order once they are ready, so they execute and write back their results in an out-of-order manner. After the instructions complete their execution, they are kept in the reorder buffer (ROB) until the commit stage, where they are guaranteed to leave the pipeline in program order.

\begin{figure}[t]
\centering
\includegraphics[width=\columnwidth]{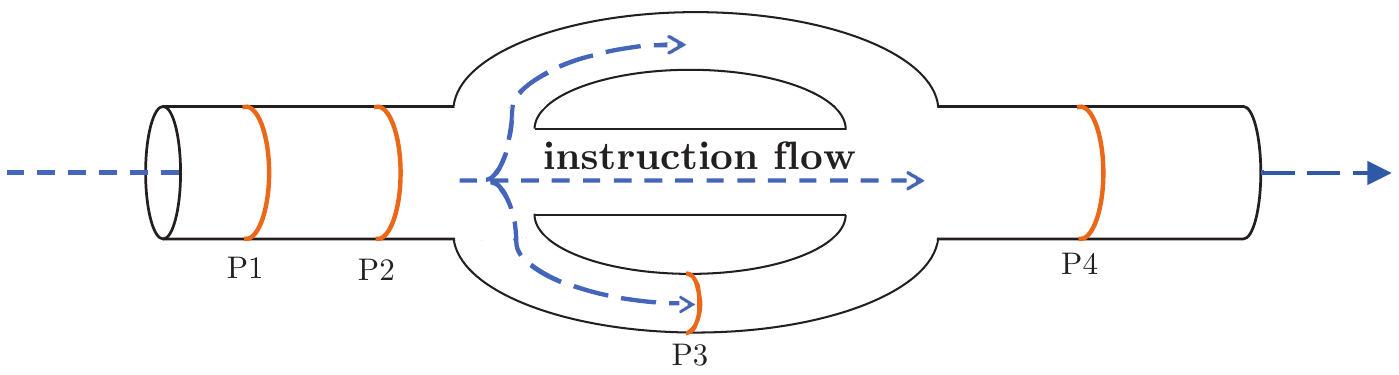}
\vspace{-0.4cm}
\caption{Possible points of the pipeline where values of the performance counters can be obtained.}
\vspace{-0.3cm}
\label{fig:dibuix-tubets} 
\end{figure}

Performance can be characterized at multiple points in the pipeline, provided that all executed instructions traverse that point. Figure \ref{fig:dibuix-tubets} illustrates this fact. In this example, the processor performance could be monitored at points P1, P2, and P4 as the input flow crosses all these points. These points could refer to the dispatch, issue, and commit stages. The general pipeline is divided into three small pipes, representing the functional units that join back to the general pipeline before instructions leave it. Notice that the overall processor performance could not be measured in one of the small pipelines (e.g., point P3) as they only observe a fraction of the overall flow. 

Previous research has evaluated performance at different points of the pipeline, showing that no point is necessarily better than another to evaluate it~\cite{multi_cpi_stacks}. 
In any case, when working on a real processor, the point (if any) to monitor the performance is constrained to the available performance counters. For instance, IBM POWER8 allows monitoring the performance at the commit stage ~\cite{symbiotic-josue}, while Intel processors do it at the issue stage ~\cite{top-down}. 
Nevertheless, the methods proposed in these works cannot be applied to ARM processors due to the widely different performance counters. Therefore, a new measurement method needs to be devised.

In order to accurately quantify the application performance, we need a handful of performance 
counters that measure the application performance at a point in the pipeline traversed by all the instructions.
Studying if this is possible and, in such a case, identifying which performance counters to use is challenging since most performance counters are not exclusive to each other. 
In fact, some of the collected events can overlap each other, which translates into their sum being below or above 100\% cycles. 
As a consequence, in practice, selecting the set of performance counters that will be used can enormously differ depending on the deployed performance counter architecture.
Therefore, to develop an accurate performance model, available performance events and their measurement capabilities must be studied stage by stage.

\subsection{Characterizing the Performance in ARM Processors: Dispatch Stalls and Dispatch Cycles}
\label{ssec:dispatch_cycles}

\begin{figure}[t]
\centering
\includegraphics[width=0.92\columnwidth]{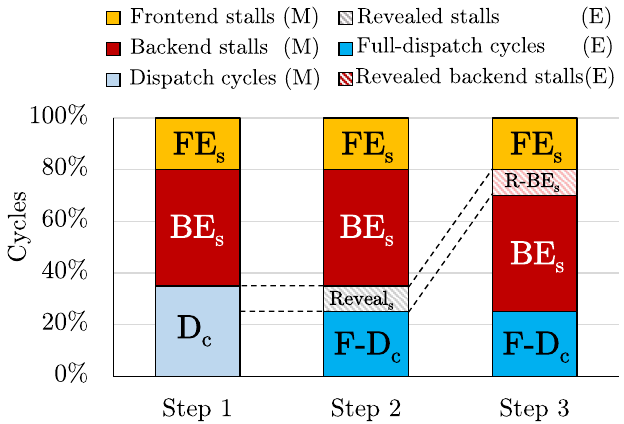}
\vspace{-0.1cm}
\caption{Characterization of cycles at the dispatch stage. Some categories are directly measured (M) with performance counters, while others are estimated (E) based on them.}
\vspace{-0.4cm}
\label{fig:dispatch-classficiation-arm} 
\end{figure}

Under the above premises, we looked into the available performance counters of ARM processors~\cite{armv8} to devise a new approach. Unlike in other architectures, we found that hardware events at the dispatch stage in ARM processors are suited to assess performance. More precisely, we found that frontend and backend stalls can be discerned at this stage with the available counters.
Moreover, compared to other processor stages, the sum of cycles measured at this stage gets closer to 100\% of the total execution cycles.
Therefore, based on these performance counters, we devised the following approach to characterize performance for the ARM processor consisting of three main steps.



\textbf{Step 1.}
We first distinguish three main events at the dispatch stage: frontend stalls, backend stalls, and no stalls, as depicted in the first column of Figure \ref{fig:dispatch-classficiation-arm}. Stalls refer to execution cycles where the dispatch is stalled. In such a case, stalls can be attributed either to the frontend or the backend. When the dispatch queue is empty, no instruction can be dispatched, and the stall should be attributed to the frontend. We refer to these stalls as \emph{frontend stalls} or $FE_s$. For example, this can occur due to stalls caused by an I-cache miss. On the contrary, when the dispatch queue is not empty but instructions cannot be dispatched due to the lack of a required resource (e.g., a ROB entry), the stall is assigned to the backend (referred to as \emph{backend stalls} or $BE_s$). These stalls are mainly due to long-latency instructions, such as a load accessing main memory, causing the ROB to block. The sum of the frontend or the backend stalls does not give the total \emph{dispatch cycles}, but the difference accounts for the cycles where at least one instruction is dispatched, represented as \emph{dispatch cycles} or $D_c$. 

To obtain the values of three events, we monitor the performance counters \emph{cpu\_cycles}, \emph{stall\_frontend}, and \emph{stall\_backend}. Table \ref{tab:performance-counters-arm} describes these hardware events along with the \emph{inst\_spec} event, whose usage will be described next.  


\textbf{Step 2.}
The events of Step 1 can be further refined. Remember that the dispatch logic can dispatch as many instructions as the dispatch width in a given cycle. This means that if the dispatch width is $N$ and a single instruction is dispatched in a cycle, that cycle will not be counted as a dispatch stall despite $N-1$ dispatch slots being wasted. In such a case, the actual stalls can be accurately estimated as $\frac{N-1}{N}\%$ for a given cycle. To consider these cases, we compute the cycles needed to dispatch the instructions, assuming that the full dispatch width (4 in our processor) is consumed. We refer to these cycles as equivalent \emph{full-dispatch cycles} ($F$-$D_c$) and calculate them as the number of executed instructions (\emph{inst\_spec} performance counter) divided by the processor dispatch width. 

The difference between $D_c$ and $F$-$D_c$ reveals a fraction of stalls due to \emph{horizontal waste} of the dispatch slots ($Reveal_s$). This fraction was hidden in the characterization of Step 1 due to performance counters limitations, which only count cycles when no instruction is dispatched. Figure~\ref{fig:dispatch-classficiation-arm} (Step 2) depicts the new characterization with the full-dispatch cycles (lower than the dispatch cycles initially measured) and the revealed stalls (small grey frame of the bar). Notice that the dispatch cycles bar of Step 1 equals the full-dispatch cycles plus the revealed stall cycles of the current Step 2.

\begin{table}[t]
\centering
\resizebox{\columnwidth}{!}{%
\begin{tabular}{|l|l|}
\hline
Counter name        & Explanation                                                                   \\ \hline
CPU\_CYCLES         & Cycles                                                                        \\ \hline
INST\_SPEC          & Operation (speculatively) executed                                              \\ \hline
\multirow{2}{*}{STALL\_FRONTEND}     & Cycles on which no operation is dispatched because \cr & there is no operation in the queue    \\ \hline
\multirow{2}{*}{STALL\_BACKEND}      & Cycles on which no operation is dispatched due to \cr & backend resources being unavailable \\ \hline
\end{tabular}
}
\vspace{0.2cm}
\caption{Hardware events gathered in the ARM processor to perform the performance characterization at dispatch.}
\vspace{-0.5cm}
\label{tab:performance-counters-arm}
\end{table}


\textbf{Step 3.}
Despite there is no performance counter available that provides a clue to attribute stalls in $Reveal_s$ (i.e., horizontal waste) to either the frontend or the backend, theoretically, most of them cannot be attributed to the frontend. Frontend stalls are mainly due to events that squash the fetched instructions (e.g., branch mispredictions) or block instruction supply (e.g., I-cache misses). In both cases, the 4 dispatch slots available in a cycle will be wasted, which will be correctly accounted for by the \emph{stall\_frontend} performance counter. In contrast, the backend can cause frequent horizontal waste, for instance, due to the required floating-point unit or cache being used but becoming available after a few cycles. Therefore, in Step 3, we assign the revealed stalls to the backend category ($R$-$BE_s$), thus resulting in just three main categories (frontend stalls, backend stalls, and full-dispatch cycles) to feed the model and estimate the synergy between pairs of applications. Different alternatives were also evaluated, such as splitting $Reveal_s$ into equal (and proportional) parts to front-end and backend stalls. We opt for the selected design choice as it is the one showing the most accurate regression model.


Finally, unlike other approaches~\cite{Eyerman_CPIacc,Eyerman_SMTacc}, we intentionally do not make any distinction between correctly executed (retired) and wrong-path (mispeculated) instructions. The devised performance classification seeks to estimate the inter-application interference at the shared dispatch slots. Therefore, from a resource consumption perspective, there is no distinction between a dispatch slot consumed by a committed instruction or by a canceled instruction. 

\section{The Synergistic Approach}

SYNPA selects the pairs of applications to be executed in each core in order to minimize inter-application interference across all the processor cores. To this end, a linear regression model with just three variables is used. This section discusses the devised regression model.

\subsection{Estimating Performance in SMT Execution}
\label{sec:model}

The model computes the performance degradation a thread suffers over isolated execution when it runs simultaneously with another thread in the same SMT core. As mentioned in Section \ref{ssec:dispatch_cycles}, the model uses three categories: full-dispatch cycles, frontend stalls, and backend stalls (which also include the horizontal waste). The reasoning behind the model is that, for example, the backend stalls of an application caused by the lack of backend resources (e.g., ROB entries) are expected to increase when contending with another application. Consequently, the sum of three categories (3rd step of Figure \ref{fig:dispatch-classficiation-arm}) gathered in SMT execution normalized to isolated execution will exceed 100\% cycles, which represents the slowdown the application suffers due to SMT execution. 

Accurately estimating to what extent each category varies is challenging as it depends not only on the characteristics of the application itself but also on those of the co-runners, i.e., their interaction within the core. To predict the variation of the categories in SMT execution, we devised a linear regression model. The regression model is based on Equation~\ref{eq:forward}, where $C^{smt}_{i,j}$ represents the value of category $C$ for application $i$ when it runs together with application $j$ on the same core, $C^{st}_{i}$ and $C^{st}_{j}$ represent the value of category $C$ for application $i$ and $j$, respectively, when they are executed in isolation in ST (single-threaded) mode, and $\alpha_{C}$, $\beta_{C}$, $\gamma{C}$, and $\rho_{C}$ are the coefficients of the model. 

\begin{equation} \label{eq:forward} 
{C}^{smt}_{i,j} = \alpha_{C} + \beta_{C}\cdot C^{st}_{i} + \gamma_{C}\cdot C^{st}_{j} + \rho_{C}\cdot C^{st}_{i}\cdot C^{st}_{j}
\end{equation}

In other words, Equation \ref{eq:forward} estimates the value for category $C$ of application $i$ when running with application $j$ on the same core as a weighted sum of the following three components:

\begin{itemize}
    \item The value of category $C$ in application $i$ when it runs in isolation ($C^{st}_{i}$).
    \item The value of category $C$ in application $j$ (the co-runner) when it runs in isolation ($C^{st}_{j}$). 
    \item The product of both values above.
\end{itemize}

Each term is multiplied by a coefficient ($\beta_{C}$, $\gamma_{C}$, and $\rho_{C}$) that weights its influence to model the category. Regarding the $\alpha_{C}$ coefficient, known as the independent term, it helps reduce the bias of the model and therefore improves its accuracy. $\alpha_{C}$, $\beta_{C}$, $\gamma_{C}$, and $\rho_{C}$ are obtained for each category as a whole, using linear regression, and therefore they do not depend on each particular application. In general, the model estimates the time for a category in SMT (${C}^{smt}_{i,j}$) as the sum of that category for the target application in isolation ($C^{st}_{i}$) plus some interference introduced by the co-runner. Notice that ${C}^{smt}_{i,j}$ and ${C}^{smt}_{j,i}$ are not the same (not symmetric) because when both applications run simultaneously on the same SMT core, the slowdown applications $i$ and $j$ suffer differs. This occurs because, within the pair, the applications have different behavior and resource demands.

\subsection{\emph{SYNPA} policy}

\begin{figure}[t]
\centering
\includegraphics[width=\columnwidth]{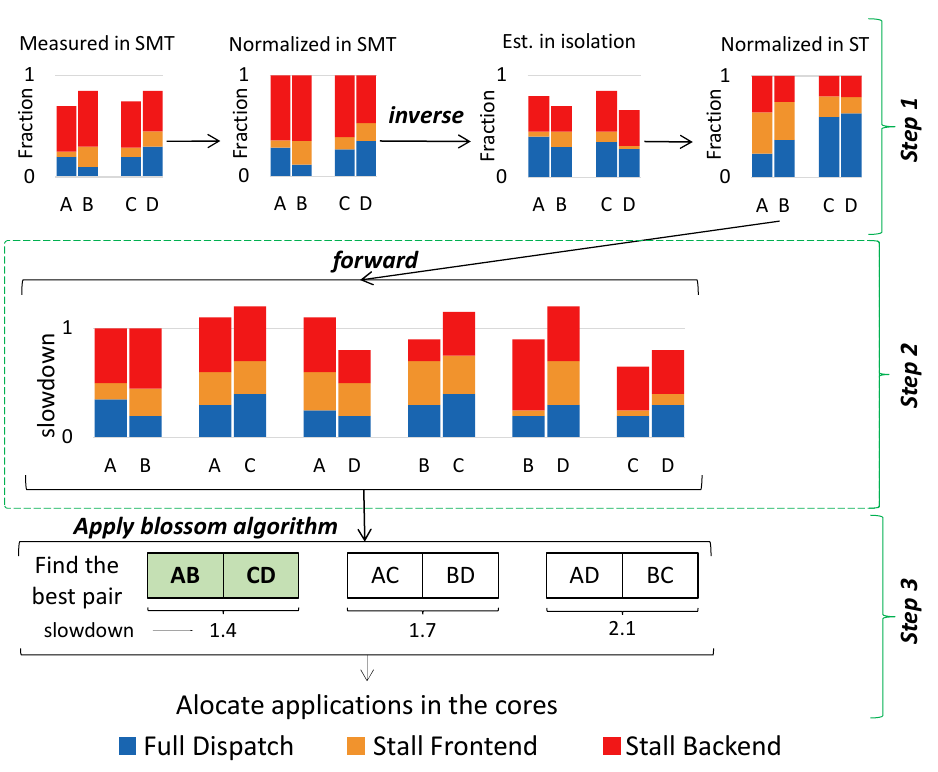}
\caption{ General overview of the steps of the SYNPA policy.} 
\vspace{-0.5cm}
\label{fig:general-scheme-model} 
\end{figure}

At each execution interval or quantum (100ms), each processor core executes a different pair of applications. Performance counters monitor the events shown in Table~\ref{tab:performance-counters-arm} on a per-application basis. 
Once each quantum expires, the policy determines which are the most synergistic pairs of applications and schedules them to maximize performance. Next, we describe the different steps that SYNPA performs, which are represented in Figure \ref{fig:general-scheme-model}. Notice that these steps are repeated each quantum until the workload execution is completed.


\textbf{Step 1.}
\textbf{Estimating the value of categories in isolation.}
As explained above, the regression model predicts the SMT category values from the ones when running in isolation (ST mode). However, during SMT execution, these ST values are not available. Fortunately, Feliu et al.~\cite{symbiotic-josue} already addressed this problem and demonstrated that the proposed interference model could be inverted to estimate, with relatively good accuracy, the $C^{st}_{i}$ and $C^{st}_{j}$ values for the categories of two applications ($i$ and $j$) running in the same SMT core. These values are computed using the values of $C^{smt}_{i,j}$ and $C^{smt}_{j,i}$, which are available in the SMT mode. Therefore, we leverage this model inversion to obtain $C^{st}_{i}$ and $C^{st}_{j}$ at runtime. To do that, after each quantum expires, SYNPA gathers the event counts for each application and computes the performance categories for the SMT execution (i.e., ${C}^{smt}_{i,j}$ and ${C}^{smt}_{j,i}$) during the last interval. These categories are normalized with respect to the execution cycles to obtain the probability of occurrence of each category event (${C'}^{smt}_{i,j}$ and ${C'}^{smt}_{j,i}$). Then, SYNPA applies the \textit{inverse model} to estimate the ST value of each category (${C}^{st}_{i}$ and ${C}^{st}_{j}$). 

\textbf{Step 2.}
\textbf{Estimating the SMT performance of application pairs.}
With ${C}^{st}_{i}$ and ${C}^{st}_{j}$, 
we can apply the regression model to predict SMT performance. The model estimates the performance degradation (i.e., slowdown) of a given application when it runs together with another application in the same SMT core. Applying the model equation twice (once for application A when it co-runs with application B and once for application B when it co-runs with application A) allows estimating the impact of A on the performance of B and vice-versa. In other words, it estimates how synergistic a pair of applications is. The synergy of all pairs of applications is estimated in this step. 

\textbf{Step 3.}
\textbf{Selecting the synergistic pairs.} 
With the synergy of each pair of applications, we can select the combination with the lowest degradation overall. However, the number of possible combinations grows quickly with the number of cores, thus increasing the overhead of selecting the optimal one by evaluating all possible combinations. To find the optimal combination with the minimum overhead, we model the selection problem as a graph problem and solve it with the Blossom algorithm~\cite{blossomalgorithm}. 
Finally, the applications are allocated to their corresponding core as determined by the most synergistic combination.


\subsection{Model Training} \label{sec:training}
Before the model can be used, we should obtain the set of coefficients $\alpha_{C}$, $\beta_{C}$, $\gamma{C}$, and $\rho_{C}$ for the regression model. As discussed earlier, we devise a model per category, not per application. This makes the proposed model flexible and able to work with any application as long as the training set is diverse enough.

To train the model, we run 80\% of the applications (22 out of 28) in isolation and create a profile with the value of the different categories and the number of committed instructions for each quantum. Next, we run in SMT mode for all the possible pairs of these training applications and collect the same data. As execution progresses slower in SMT mode, the number of committed instructions allows us to map the category values of an application when it runs in isolation to the corresponding values when it runs in SMT mode with another application.

A random subset of the execution quanta was selected to build the model. The model of each category was trained using the data of these quanta, and the coefficients were selected, minimizing the model error as much as possible. As long as the set of applications used for training is diverse enough, the model only needs to be trained once. After that, the model coefficients can be used to select the most synergistic pairs of applications in the system. Therefore, the cost of training the model can be quickly amortized thanks to the performance benefits it provides.

Note, however, that the model was built for compute-intensive scientific workloads, like those in the SPEC CPU suite, so it is valid for workloads showing this behavior. In order to cover applications showing distinct behavior (e.g., graph workloads), the model needs to be re-trained with the new applications.

\section{Experimental Framework}
This section describes the experimental framework, which consists of the experimental machine and the manager developed to carry out the experiments, the workload design, and the evaluation methodology.

\subsection{Platform: System and Manager}
SYNPA approach has been implemented in a Cavium ThunderX2 CN9975 processor~\cite{ThunderX2} and a 64GB DRAM main memory. This processor is based on the Vulcan microarchitecture \cite{vulcan} and implements the ARMv8.1A \cite{armv8} instruction set. It is composed of 28 SMT4 cores and has a 28MB shared LLC. Despite the processor supporting up to 4 threads, it was configured in the BIOS as SMT2 (56 SMT2 cores) since HPC workloads are core resource-hungry and SMT4 can severely damage the performance mainly due to small (32KB) L1 data caches, which make these applications experience a high number of misses, so yielding the system to poor performance. This is the main reason why Intel mostly implements SMT2 processors even with larger 48KB caches. 
Table \ref{tab:platform} summarizes the main features of the system, including specific parameters of the core microarchitecture. The system runs a CentOS Linux 7 (AltArch) distribution with kernel 4.18.

\begin{table}
\centering
\begin{tabular}{ |l|l| }
\hline
\multicolumn{2}{|c|}{Cavium
ThunderX2 CN9975 processor} \\\hline
 \# Cores & 28 \\ 
 \# Threads & 112 \\\hline
\multicolumn{2}{|c|}{Core microarchitecture} \\\hline 
  Dispatch width & 4 \\
 ROB size & 128 entries \\
 IQ size & 60 entries \\
 Load/Store buffer & 64/36 entries \\
 \# Issue ports & 6 \\\hline
 \multicolumn{2}{|c|}{Memory subsystem} \\\hline 
 L1I/L1D & 32 KB / 32 KB \\
 L2 & 256 KB \\
 Shared LLC & 28 MB \\
 Main memory & 64 GB \\\hline
\end{tabular}
\vspace{0.3cm}
\caption{Main features of the experimental processor and memory subsystem.\label{tab:platform}}
\vspace{-0.7cm}
\end{table}

To carry out the experiments, we prototype SYNPA as a user-level thread manager that gathers the required performance counters, uses them to estimate the synergy of the possible pairs of applications, selects the optimal combination, and performs the corresponding thread-to-core mapping. The developed manager uses the \emph{perf} tool to configure and read performance counters and controls the thread execution and allocation to the different cores using the \textit{sched\_setaffinity} system call. To fairly assess the performance of the Linux scheduling policy, we use the same thread manager but allow Linux to select which thread runs in each core (i.e., the thread-to-core allocation).

\subsection{Measurement Methodology}
\label{sec:workloads} 

To evaluate SYNPA, we have designed a wide set of 8-application HPC workloads. 
These workloads combine workloads used for training the model with other workloads that were reserved to evaluate it with \emph{new} applications.

As we have already discussed, the interference in the shared resources and, therefore, the performance of an SMT processor strongly varies according to the applications executed concurrently on each core. In order to design insightful workloads and evaluate the performance gains that \textit{SYNPA} is able to achieve in different scenarios, we first characterized the 28 applications studied in this paper. 

\begin{figure}[t]
\centering
\includegraphics[width=\columnwidth]{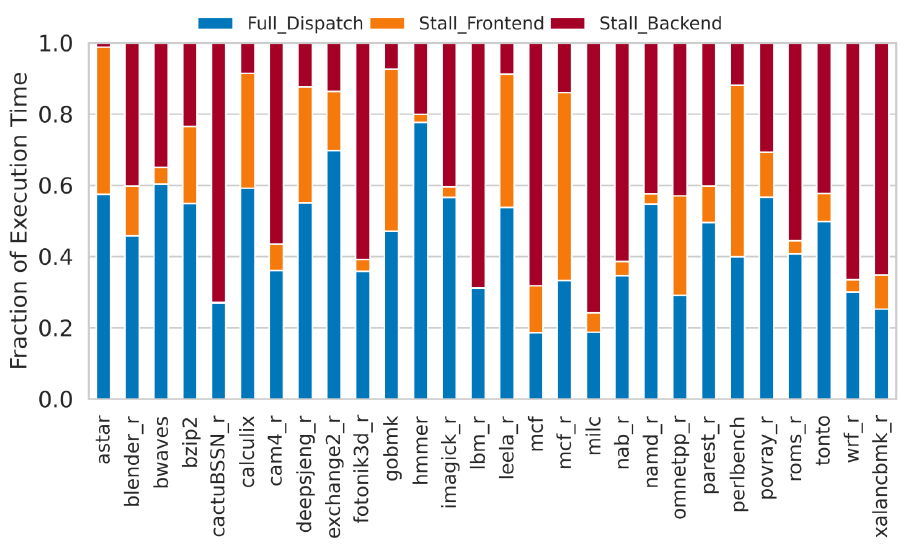}
\caption{ Characterization of the applications in isolated execution. }
\label{fig:characterization-isolated} 
\end{figure}

Figure \ref{fig:characterization-isolated} shows the execution time distribution for each application considering the dispatch stalls, both at the backend and frontend, and the full-dispatch cycles when running each application in isolation. Considering this characterization, we classify the studied applications into three main groups depending on the fraction of backend and frontend stalls. The first group, referred to as \textit{Backend bound}, includes the applications whose dispatch stalls due to the backend represent more than $65\%$ of the cycles. The second group, \textit{Frontend bound}, includes the applications whose frontend dispatch stalls represent more than $35\%$. Finally, the third group, (\textit{Others}), consists of the remaining applications and includes applications with different behaviors whose full dispatch cycles range from 20\% (\emph{hmmer}) to 61.4\% (\emph{nab\_r}). Table \ref{tab:applications_x_group} shows the classification of each benchmark in these three groups.

\begin{table}[t]
\resizebox{\columnwidth}{!}{%
\begin{tabular}{|c|c|l|}
\hline
Group & Dominant Category                  & Applications \\ \hline
\multirow{2}{*}{Backend bound }
& \multirow{2}{*}{ Backend stalls $>65\%$  } & cactuBSSN\_r, lbm\_r, mcf,\cr & & milc, xalancbmk\_r,wrf\_r \\\hline
 \multirow{2}{*}{Frontend bound }& \multirow{2}{*}{ Frontend stalls $>35\%$} & astar,  gobmk, leela\_r, \cr & & mcf\_r, perlbench \\ \hline

\multirow{5}{*}{Others }        &  \multirow{5}{*}{rest of apps}   &  blender\_r, bwaves, bzip2, calculix,  \cr & & cam4\_r, deepsjeng\_r, exchange2\_r,  \cr & & fotonik3d\_r, hmmer, imagick\_r, nab\_r, \cr & & namd\_r, omnetpp\_r,  parest\_r, povray\_r, \cr & & roms\_r, tonto \\ \hline

\end{tabular}%
}
\vspace{0.2cm}
\caption{ 
Benchmark grouped backend and frontend bound according to their fraction of backend and frontend dispatch stalls, respectively.
}
\vspace{-0.5cm}
\label{tab:applications_x_group}
\end{table}


As we aim to evaluate a representative range of scenarios, we built backend-intensive workloads, frontend-intensive workloads, and mixed workloads to evaluate respective scenarios mainly dominated by backend-intensive applications, frontend-intensive applications, and a mix of them. Each workload consists of 8 applications. Backend-intensive workloads include 5 or 6 applications randomly selected from the backend bound group and the remaining applications randomly selected from the \emph{Others} group. Frontend-intensive workloads are built analogously but take most applications from the frontend bound group. Finally, mixed workloads are built randomly, selecting half of the applications from the backend bound group and the other half from the frontend bound group. We evaluate a total amount of 20 workloads: 5 backend-intensive \texttt{(be0-be4}), 5 frontend-intensive \texttt{(fe0-fe4}), and 10 mixed (\texttt{fb0-fb9}) workloads.

Individual applications of the workloads take different times to execute. Therefore, to keep the number of applications constant along the workload execution, we used the following methodology. First, we executed each application in isolation for 60 seconds and recorded its number of retired instructions. This number will be the target number of instructions running in the workload. When running multi-program workloads, we record the performance of each individual application at the time it reaches the target number of instructions. Then, the application is relaunched to keep the workload constant across the entire experiment. The workload execution completes when the slowest application reaches its target number of instructions. 

Finally, experimental results presented for each workload in the next section were obtained, executing nine times each workload. We obtained the average of all these executions and calculated the variation of each execution. If an execution shows an excessive variation (90 approx. difference between the value of one execution and the average value of all the executions), it is discarded in order to obtain a variation coefficient of less than 5\%. Notice that we follow this methodology to avoid those values that are not representative of the real behavior of the workloads.


\section{Experimental Evaluation}

This section presents the coefficient values and accuracy of the devised model. After that, the performance of SYNPA is evaluated, in terms of turnaround time, fairness, and IPC. SYNPA performance is compared to the Completely Fair Scheduler of the CentOS Linux 7 (AltArch). It is important to remark that the SYNPA approach cannot be compared to existing approaches for Intel and IBM processors, as the events they use are not available in the ARM processor.

\subsection{Coefficient Values and Model Accuracy}

\begin{table}[t]
\centering
\begin{tabular}{|c|r|r|r|r|}
\hline
Category & \multicolumn{1}{c|}{$\alpha$} & \multicolumn{1}{c|}{$\beta$} & \multicolumn{1}{c|}{$\gamma$} & \multicolumn{1}{c|}{$\rho$}  \\ \hline
Full-dispatch cycles & 0.0072  & 0.9060  & 0.0044  & 0.0314 \\ \hline 
Frontend stalls            & 0.2376  & 1.4111  & 0  & 0  \\ \hline 
Backend stalls             & 0.2069  & 0.3431  & 1.4391  & 0   \\ \hline 
\end{tabular}%
\vspace{0.2cm}
\caption{ Model coefficients for the three categories.}
\vspace{-0.5cm}
\label{tab:parameters-linear-regression}
\end{table}

We built the linear regression model discussed in Section \ref{sec:model} using the training method described in Section \ref{sec:training}.
Table \ref{tab:parameters-linear-regression} presents the coefficient values obtained with linear regression to model the behavior of each category (see Section \ref{sec:training}). As expected, the backend stalls category is the most affected by interference, 
since threads share (and compete for) critical backend resources such as the data cache, which can significantly increase SMT execution time, as experimental results will show. Regarding the frontend stalls category, it grows in SMT execution proportionally to isolation, even though a relatively high $\alpha_{C}$ coefficient is added. 
To some extent, this model reflects the fact that the frontend stages are limited due to the IFetch policies, which only allow a single thread to access the ICache at a given processor cycle. This fact can significantly affect frontend time in SMT mode. Note that this category mainly depends on the application itself (high $\beta_{C}$, and low $\gamma_{C}$ and $\rho_{C}$). 

Regarding the number of dispatch cycles in SMT, it can be observed that the $\beta_{C}$ value is lower than in isolation. The main reason is that in SMT execution, an application makes slower progress than in ST mode. The number of stalls grows and, therefore, the percentage of full-dispatch cycles reduces. For this category, the $\rho_{C}$ coefficient is also not negligible, which indicates that the dispatch rate in SMT mode is affected by the dispatch rate of both threads.

To assess the accuracy of the model, we obtained the Mean Square Error (MSE) for the backend stalls, frontend stalls, and full-dispatch cycles categories, which are $0.1583$, $0.0703$, and $0.0021$, respectively. The error in the backend stalls category is the highest because this category is the most sensitive to interference variations as it comprises several critical backend resources whose contention is highly dependent on the behavior of the co-runner (as the value of $\gamma_{C}$ corroborates).

At first glance, one might think that the more categories a model includes, the higher its accuracy and performance. More precisely, the higher would be the quality of the synergistic pairs that would yield the highest system performance.
 However, this is not necessarily the case. Notice that the more categories, the more sources of inaccuracy due to model deviations and the higher the model overhead. 
In fact, we initially developed a preliminary model with ten categories and achieved worse results. More precisely, the backend category was initially split into seven categories depending on the components that raise the dispatch stalls (e.g., ROB full, IQ full, etc.). After developing a simpler model, we found that the sum of the error deviations with more components exceeds the errors of only considering the backend category as a single category. This means that a three-category model presents a big advantage, not only because of its simplicity but also because it presents lower overhead due to fewer equations being used.

\subsection{ Turnaround Time Analysis}

\begin{figure}[t]
\centering
\includegraphics[width=\linewidth]{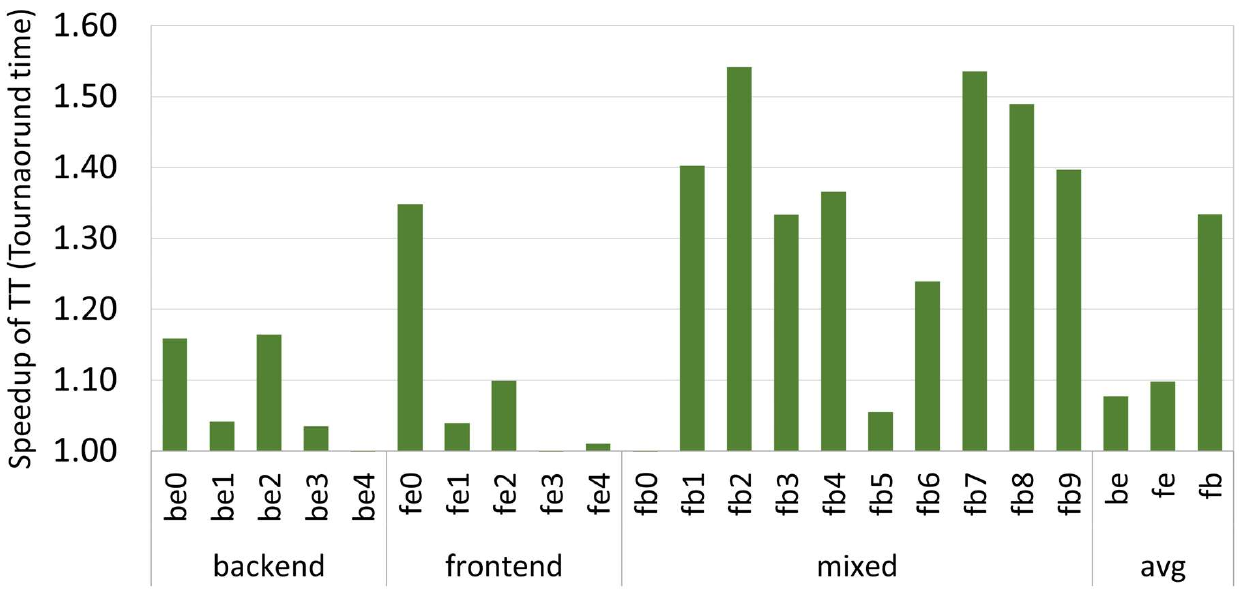}%
\caption{
Speedup of the TT over Linux.}
\vspace{-0.5cm}
\label{fig:speedup} 
\end{figure}

The turnaround time (TT) of a workload is defined by the time the workload takes to execute. More precisely, by the time taken by the slowest benchmark of the workload. Figure \ref{fig:speedup} presents the speedup of the TT of SYNPA over Linux for the studied workloads and the average speedup for each workload type. As observed, SYNPA achieves significant improvements over Linux. Focusing first on backend-intensive workloads (\texttt{be0-be4}), it can be appreciated that SYNPA performs much better than Linux, reaching an average speedup of around 18\%. The benefit diminishes significantly in the frontend-intensive workloads (\texttt{fe0-fe4}), despite SYNPA still outperforming Linux by approximately 8\%. This indicates that there is little margin for the thread allocation policy to improve performance in workloads dominated by frontend bound benchmarks. This seems rather intuitive as the performance of these benchmarks already suffers significantly from ICache misses in stand-alone execution. Thus, this problem is likely to aggravate even more in SMT execution. An interesting observation is that SYNPA performance improvements significantly rise with mixed workloads (\texttt{fb0-fb9}), where backend bound applications are mixed with frontend bound applications. In these workloads, SYNPA is able to improve the TT up to 1.55 in \texttt{fb2}, i.e., 55\% compared to Linux, with an average speedup of 36\%. This means that SYNPA is able to find, at runtime, synergistic pairs of applications whose behavior complements each other and allocates them in the same SMT cores to maximize the overall performance.

\begin{figure*}[ht]
\centering
    \begin{subfigure}[t]{1\columnwidth}
    \centering
    \includegraphics[width=0.65\textwidth]{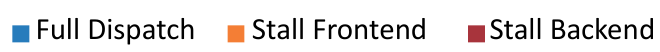}%
    \end{subfigure}
    \hfill
    
    \begin{subfigure}[t]{0.3\textwidth}
    \centering
    \includegraphics[width=1.05\linewidth]{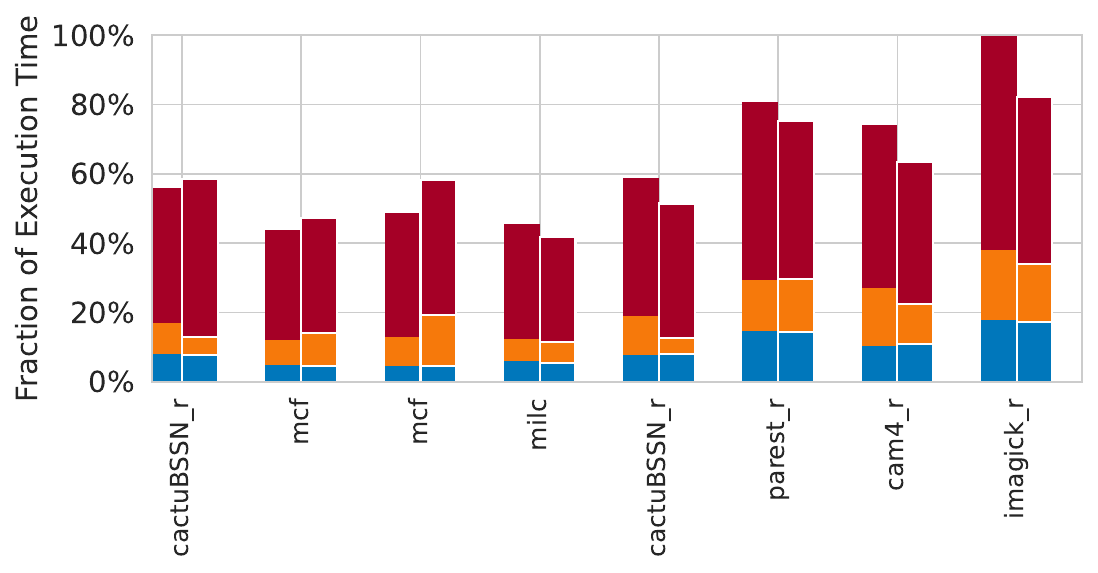}%
    \caption{Workload \texttt{be1}}       
    \label{subfig:wk1}
    \end{subfigure}
    ~
    \begin{subfigure}[t]{0.3\textwidth}
    \centering
    \includegraphics[width=1.05\linewidth, height =0.551\linewidth]{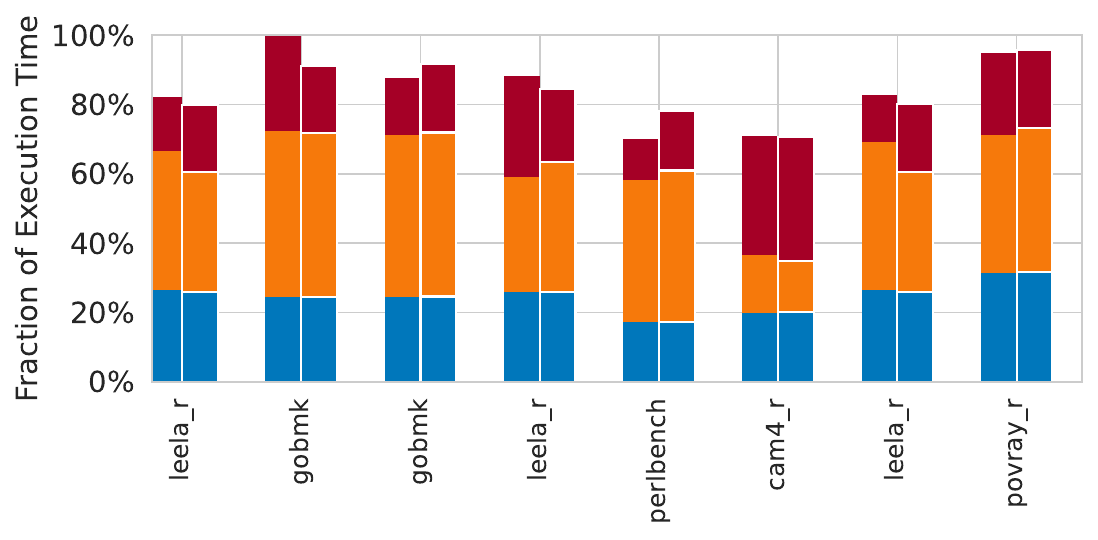}
    \caption{Workload \texttt{fe2}}     
    \label{subfig:wk7}
    \end{subfigure}
     ~
    \begin{subfigure}[t]{0.3\textwidth}
    \centering
    \includegraphics[width=1.05\linewidth]{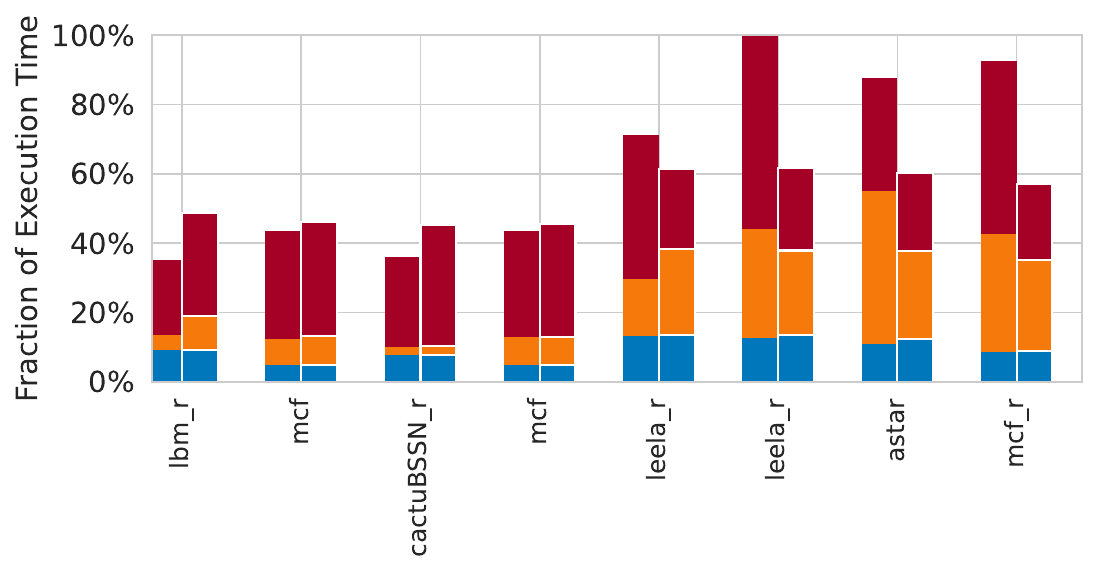}%
    \caption{Workload \texttt{fb2} }
    \label{subfig:wk12}
    \end{subfigure}

    \caption{Characterization for the 8 applications of three workloads. The left bar of each application corresponds to the Linux scheduler, and the right bar corresponds to the SYNPA policy. 
    }
    \label{fig:carac_x_mix}
\end{figure*}

To provide insights that explain the reason why SYNPA is able to achieve so high TT improvements, we measured the values of the three main categories used by the model for each individual benchmark when running the entire workload in SMT execution. For illustrative purposes, Figure \ref{fig:carac_x_mix} illustrates the behavior of three workloads: one backend-intensive (\texttt{be1}), one frontend-intensive (\texttt{fe2}), and one mixed workload (\texttt{fb2}). For each workload, the corresponding figure shows the characterization of each of the eight applications that compose the workload when running with Linux (left bar) and when using SYNPA (right bar). The execution time of each application is normalized to the slowest application of the workload. 

It can be observed that in the frontend-intensive workload (\texttt{fe2}, Figure \ref{subfig:wk7}), the \texttt{stall\_backend} category is really low. Therefore, SYNPA has little chance of reducing these stalls. In addition, the frontend fraction is significantly high across all the applications of the workload. Consequently, the proposed policy has little room to improve performance as the applications that form the workload are not complementary. In contrast, in the backend-intensive workload  (\texttt{be1}, Figure \ref{subfig:wk1}), SYNPA can significantly reduce the \texttt{stall\_backend}, improving overall performance. This happens mainly because the \texttt{stall\_frontend} category rises and drops at run-time due to the application's phase behavior along their execution time. Therefore, in this case, there is room for SYNPA to improve performance. Finally, the results in the mixed workload \texttt{fb2}, Figure \ref{subfig:wk12} are impressive. In this workload, it can be observed that SYNPA significantly reduces both the \texttt{stall\_backend} and \texttt{stall\_frontend}. The average values of each category (Figure \ref{fig:speedup}) support these conclusions since the best average speedup belongs to the mixed group, followed by the backend-intensive group, and finally, the frontend-intensive group.



\subsection{Workload Example Analysis}
\begin{figure*}[ht]
\centering
    \begin{subfigure}[t]{1\columnwidth}
    \centering
    \includegraphics[width=0.65\textwidth]{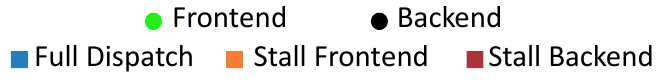}%
    \end{subfigure}
    \hfill
    
    \begin{subfigure}[t]{0.9\columnwidth}
    \centering
    \includegraphics[width=0.8\linewidth]{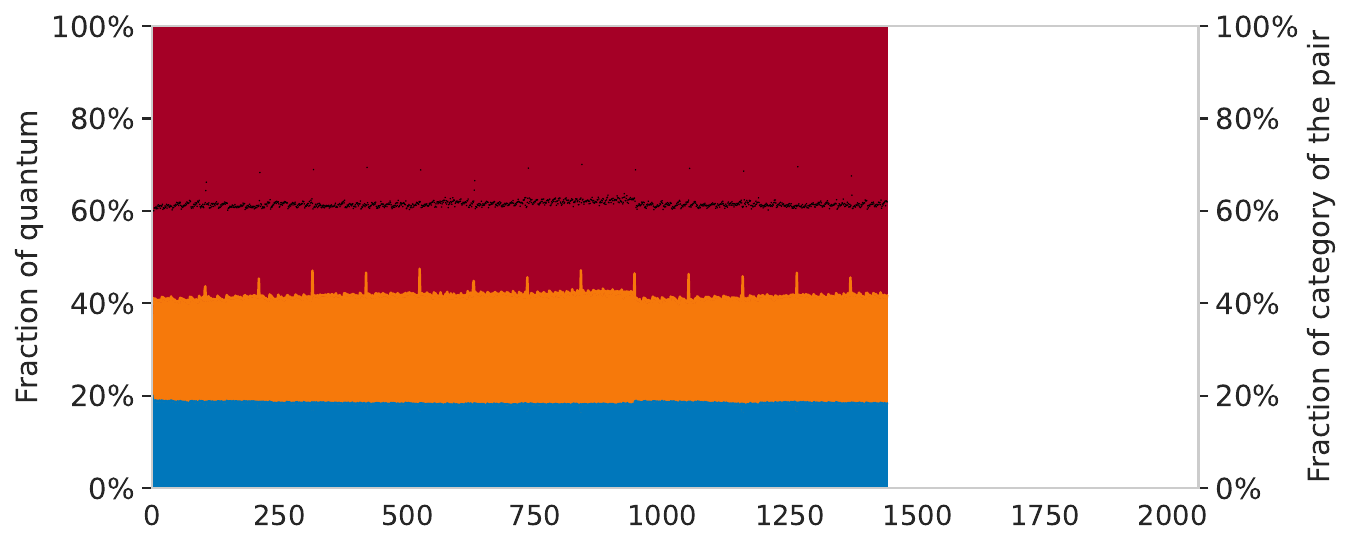}%
    \caption{Application leela\_r (04) with Linux}       
    \label{subfig:04linux}
    \end{subfigure}
    ~
    \hspace{1cm}
    \begin{subfigure}[t]{0.9\columnwidth}
    \centering
    \includegraphics[width=0.8\linewidth]{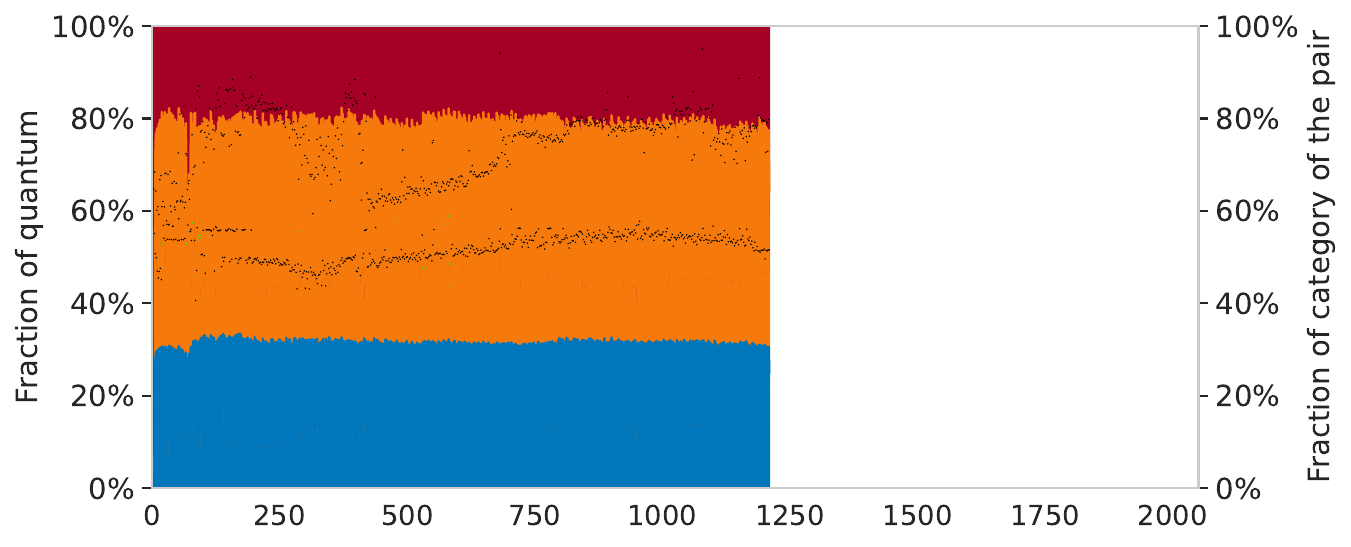}%
    \caption{Application leela\_r (04) with SYNPA}     
    \label{subfig:04synpa}
    \end{subfigure}
    \hfill
    \begin{subfigure}[t]{0.9\columnwidth}
    \centering
    \includegraphics[width=0.8\linewidth]{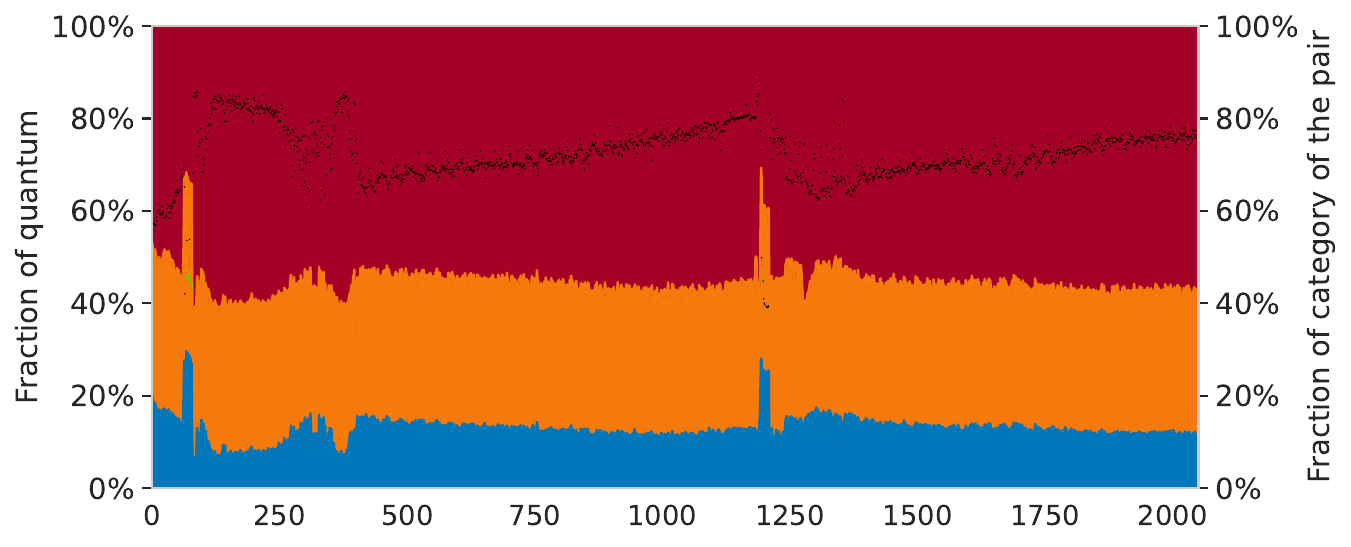}%
    \caption{Application leela\_r (05) with Linux}
    \label{subfig:05linux}
    \end{subfigure}
    ~
    \hspace{1cm}
    \begin{subfigure}[t]{0.9\columnwidth}
    \centering
    \includegraphics[width=0.8\linewidth]{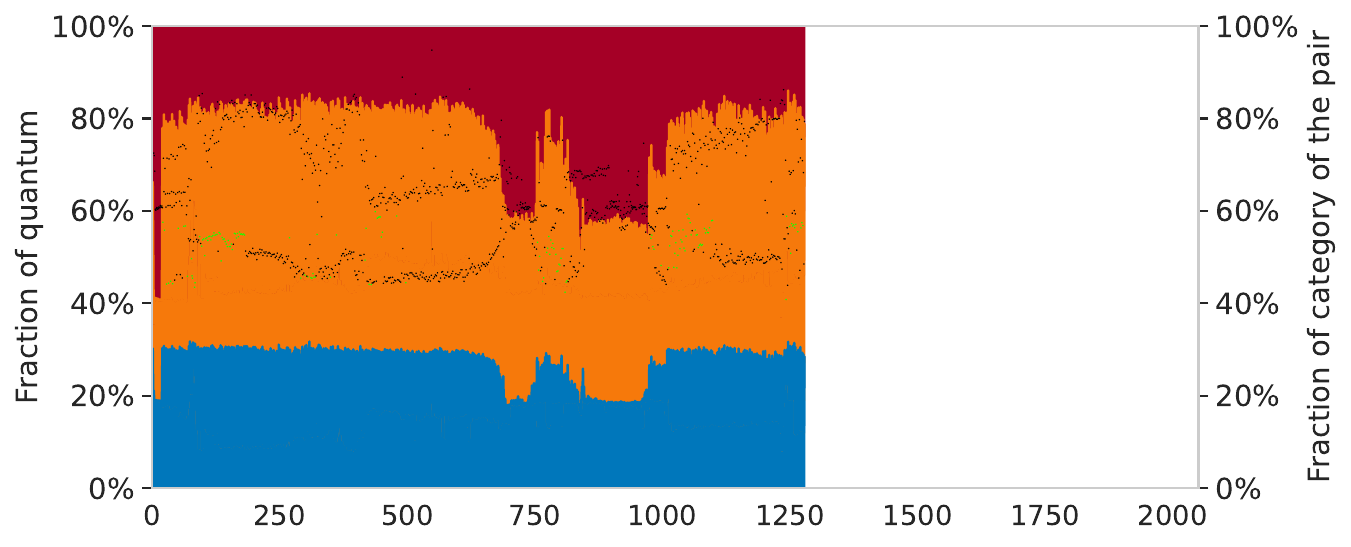}%
    \caption{Application leela\_r (05) with SYNPA}     
    \label{subfig:05synpa}
    \end{subfigure}
    \caption{ Dynamic characterization of an execution for the two applications leela\_r of \texttt{fb2} with both policies (Linux or SYNPA) 
    }
    \label{fig:carac_leela_din}
    \vspace{-0.25cm}
\end{figure*}

\begin{table}[t]
\resizebox{\columnwidth}{!}{%
\begin{tabular}{|r|c|c|c|c|c|c|c|c||c|}
\hline
pairs \%     & 
\rotatebox{90}{lbm\_r} & \rotatebox{90}{mcf}   & \rotatebox{90}{cactuBSSN\_r} & \rotatebox{90}{mcf}   & \rotatebox{90}{leela\_r} & \rotatebox{90}{leela\_r} & \rotatebox{90}{astar} & \rotatebox{90}{mcf\_r}& \rotatebox{90}{diff. group}
\\ \hline
\multirow{2}{*}{lbm\_r} & 0.00  & 0.00 & 0.00 & 0.00  & 0.00  & 0.00  & 0.00 & 0.00  & 0.0\% \\ 
& 0.00 & 3.68 & 60.89 & 11.83  & 1.80 &  14.89 & 4.55 & 2.43 & 23.7\% \\ \hline
\multirow{2}{*}{mcf}& 0.55 & 0.00 & 0.39  & 0.00 & 0.00 & 0.24  & 0.08 & 0.08 & 70.2\% \\   
& 3.13 &  0.00 & 5.96  &11.52 & 18.42 & 6.66  & 26.65 & 26.41 & 79.1\%\\ \hline
\multirow{2}{*}{cactu} & 0.00 & 0.00& 0.00& 0.00  & 0.00 & 0.00 & 0.00 & 0.00 & 0.0\% \\ 
& 60.89 & 6.35 & 0.00 & 3.76 & 3.29 & 0.94 & 5.33 & 19.51 & 29.1\% \\ \hline
\multirow{2}{*}{mcf}  & 0.24 & 0.08 & 0.24& 0.00 & 0.00 & 0.31 & 0.24&  0.00 & 50.4\% \\ 
& 11.60 & 11.44 & 3.53 & 0.00 & 27.74 & 27.82 & 16.54 & 0.31 & 73.2\% \\ \hline
\multirow{2}{*}{leela\_r}    & {\cellcolor{verd}1.80} & {\cellcolor{verd}17.79}  & {\cellcolor{verd}3.29} & {\cellcolor{verd}27.19} 
& {\cellcolor{yellow}0.00} & {\cellcolor{yellow}0.00} & {\cellcolor{yellow}2.12} & {\cellcolor{yellow}0.24} & 95.5\% \\

& {\cellcolor{yellow}0.00} & {\cellcolor{yellow}0.63}& {\cellcolor{yellow}0.00} & {\cellcolor{yellow}0.55} 
& {\cellcolor{verd}0.00}   &{\cellcolor{verd}0.00} & {\cellcolor{verd}44.59} & {\cellcolor{verd}1.88} &  97.5\%\\ \hline

\multirow{2}{*}{leela\_r}    & {\cellcolor{verd}5.25}   & {\cellcolor{verd}6.90}  & {\cellcolor{verd}0.94} & {\cellcolor{verd}27.74} 
& {\cellcolor{yellow}0.00} & {\cellcolor{yellow}0.00}  & {\cellcolor{yellow}0.00}& {\cellcolor{yellow}1.18} & 82.8\% \\ 
& {\cellcolor{yellow}9.64} & {\cellcolor{yellow}0.00} & {\cellcolor{yellow}0.00} & {\cellcolor{yellow}0.39} &{\cellcolor{verd}0.00}& {\cellcolor{verd}0.00}& {\cellcolor{verd}0.00}& {\cellcolor{verd}48.04} & 82.7\% \\ \hline
\multirow{2}{*}{astar}   &  4.55 &  26.41 &  3.45 & 16.54 & 1.41 &  0.00 & 0.00 & 0.00 & 97.3\% \\ 
&  0.00 & 0.31 & 1.88 & 0.24 &  45.30 & 0.00  &0.00 & 0.00  & 95.0\% \\ \hline
\multirow{2}{*}{mcf\_r}  & 2.43 &  26.18 & 11.99 & 0.00 &  0.24 & 12.23 & 0.00 & 0.00 & 76.5\% \\ 
&  0.00 & 0.31 & 7.52 & 0.31 & 1.88 & 36.99 & 0.00 & 0.00 &  82.7\%\\ \hline
\end{tabular}%
}
\vspace{0.2cm}
\caption{Percentages of pairs in workload \texttt{fb2} with SYNPA. The number at the top of each cell corresponds to the percentage where the application is \textit{frontend}, and the number at the bottom corresponds to the \textit{backend}.
}
\vspace{-0.7cm}
\label{tab:percentages-pairs}
\end{table}

Let us explain in detail through a workload example how SYNPA produces synergy among applications, collocating them in the same SMT core, while Linux does not. Let us take as an example the workload \texttt{fb2}. First, notice that the Linux OS scheduler is unaware of the thread's behaviors \cite{cal-marta}; that is, the Linux scheduler 
assigns applications to cores without considering their behavior.
In the case of  \texttt{fb2}, the pairs made by Linux are:\texttt{
(lbm\_r(\textit{00}), leela\_r(\textit{04})), 
(mcf(\textit{01}), leela\_r(\textit{05})), 
(cactuBSSN\_r(\textit{02}), astar(\textit{06})),
and (mcf(\textit{03}), mcf\_r(\textit{07}))}. 
The order of each application within the workload as the policy reads it is indicated in brackets. Once allocated, an application remains in the core until its execution finishes.
Looking at Figure \ref{fig:carac_x_mix}, it can be seen that one of the two \texttt{leela\_r} applications is the critical application that defines the TT (i.e., the slowest one), mainly because the \texttt{stall\_backend} category significantly rises. The other instance presents a distinct behavior as it has a different co-runner. Consequently, despite being two instances of the same application, one of them experiences around a 15\% higher slowdown than the other, which can be considered a significant drawback of Linux. On the contrary, it can be observed that both instances of the application present roughly the same performance with the SYNPA policy.

Let us study the way SYNPA behaves in this workload. Each cell in Table \ref{tab:percentages-pairs} shows the percentage of time each pair (row and column) of applications is selected 
by SYNPA. Each cell \emph{(X,Y)} includes two numbers. The number at the top of each cell represents the number of times (in percentage) that application \emph{X} is classified as frontend when paired with application \emph{Y}.
The bottom number quantifies the fraction of time that application \emph{X} was classified as backend when being paired with application \emph{Y}. 
Thus, the sum of the values of the cells in a row (except the last cell) is 100\%.
For instance, in the cell (\texttt{astar}, \texttt{cactuBSSN\_r}), 
\texttt{astar} behaves 
3.45\% of times as frontend and 1.88\% as a backend, being scheduled together with \texttt{cactuBSSN\_r} 5.33\% of times.
Looking at the cell (\texttt{cactuBSSN\_r}, \texttt{astar}) it can be appreciated that 
\texttt{cactuBSSN\_r} behaves as a backend application for the entire execution when scheduled with \texttt{astar}.

In the colored rows, we study the behavior of the two instances of \texttt{leela\_r} that are included in the mix.
Each of the two instances of \texttt{leela\_r}, an application categorized as frontend bound application (see Table \ref{tab:applications_x_group}), is paired with an application categorized as backend bound (\texttt{lbm\_r}, \texttt{mcf}, and \texttt{cactuBSSN\_r}) for more than half of the execution time, represented by the four first cells of the row, while the following four cells give the figures where \texttt{leela\_r} is paired with a frontend bound application (either the other instance of \texttt{leela\_r} or \texttt{astar} and \texttt{mcf\_r}).

Although \texttt{leela\_r} is categorized as frontend bound, at execution, it can exhibit frontend (numbers at the top of each cell) and backend (bottom numbers) behaviors.
Numbers highlighted in green correspond to \emph{synergistic pairs}, where interference is minimized (i.e., the best decision).
For instance, when the first instance of \texttt{leela\_r} has a frontend behavior (top of the cells), the values colored in green account for the times when it is paired with an application categorized as backend bound.
Conversely, the green values 
represent times when \texttt{leela\_r} behaves as backend, and it is paired with applications categorized as frontend bound.

For each row, the last column (labeled as \emph{diff. group}) gives the percentage of synergistic pairs, which is computed as the quotient of the sum of the numbers (top or bottom of each row) colored in green to the sum of the total row.
In total,
in $95.5$\% of the intervals where the first instance of \texttt{leela\_r} has a frontend behavior, it is paired with an application categorized as backend bound.
This means that only in $4.5$\%
of the intervals where it has a frontend behavior, it is paired with a frontend bound application (numbers highlighted in yellow), which is not the optimal decision. However, in some intervals, it might be impossible to pair all applications with co-runners from a different category.

\begin{figure}[t]
\centering
\includegraphics[width=\columnwidth]{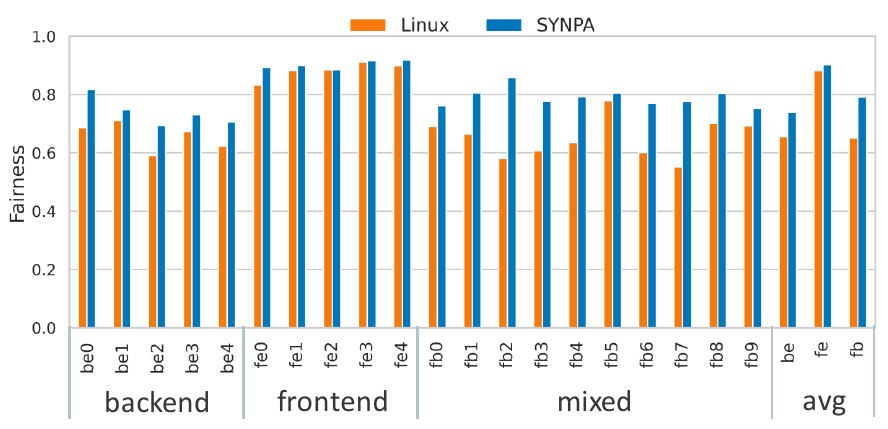}
\caption{Fairness comparison of Linux and SYNPA.}
\vspace{-0.5cm}
\label{fig:unfairness} 
\end{figure}

Finally, to analyze in detail the runtime behavior of the two instances of \texttt{leela\_r} when applying both allocation policies (Linux and SYNPA), we show the dynamical behavior of these applications during their execution in Figure \ref{fig:carac_leela_din}. This figure represents the dynamic characterization of the \texttt{leela\_r} applications (full dispatch, stall frontend, and stall backend), their turnaround time, and the value of the predominant category of the co-runner application in each quantum (either frontend or backend; dotted line in the figure). For instance, Figure~\ref{subfig:04synpa} represents the execution of the first instance of \texttt{leela\_r} applying SYNPA. In this case, it can be seen that the turnaround time of the application is around 1230 intervals. The dynamic characterization is presented in a stacked area, and in this execution \texttt{leela\_r} presents a 20\% of stalls of the backend along the entire execution. As we can see, compared with Figure~\ref{subfig:04linux}, SYNPA almost doubles the value of full dispatch cycles and decreases by one-third the value of the backend stalls, thus performing better with SYNPA.

In addition, the figure also represents the predominant category and the value of this category for the co-runner of every quantum of execution (\textit{Category fraction of the pair}). This information is presented with a point per interval. These results show that the achieved improvements result from finding the best pair for the application in each quantum based on the behavior of all the applications of the workload. The best decision is to pair applications with complementary behavior to minimize interference for the resources. SYNPA follows this criterion unless it is impossible to apply due to the behavior of the workload applications in a particular quantum. Linux always pairs \texttt{leela\_r} with a high-backend application even when  \texttt{leela\_r} is mainly limited by the backend when applying this decision. This scenario is common for the two instances of \texttt{leela\_r} with the Linux scheduler. Choosing the co-runner based on the behavior of all the applications improves the execution of both \texttt{leela\_r} in mainly three ways: a lower turnaround time (TT), a lower fraction of the backend stalls category, and a higher fraction of the full dispatch category.

\subsection{Fairness and IPC Analysis}
Running the most synergistic pairs of applications in the SMT cores reduces the number of stalls, leading not only to shorter turnaround times but also to higher system fairness and throughput. This section evaluates the benefits that SYNPA achieves compared to Linux in terms of fairness and IPC.

Fairness \cite{fairness} is evaluated for a given workload as one minus the standard deviation ($\sigma$) of the individual speedups (IS) divided by the average ($\mu$) of the IS. 
The individual speedup of each application is the ratio between the IPC it obtained in SMT execution (with SYNPA and Linux, respectively) and its IPC in isolated execution. A fairness value of 1 means that the system is completely fair.

Figure \ref{fig:unfairness} shows the fairness of the Linux and SYNPA approaches. 
As we can see, SYNPA achieves significantly higher fairness than Linux in the mixed workloads and higher fairness in the backend-intensive ones. Only in the frontend-intensive workloads the fairness difference between SYNPA and Linux is small, even though SYNPA still performs fairer than Linux. In these workloads, we also observe the maximum fairness since all applications in the workloads make slow progress. These fairness results were expected as SYNPA is able to reduce the increase of stalls due to inter-application interference, allowing the applications to progress more homogeneously. The fairness differences are up to about 48\% in workload fb2 and, on average, are about 25\%. 

\begin{figure}[t]
\centering
\includegraphics[width=\columnwidth]{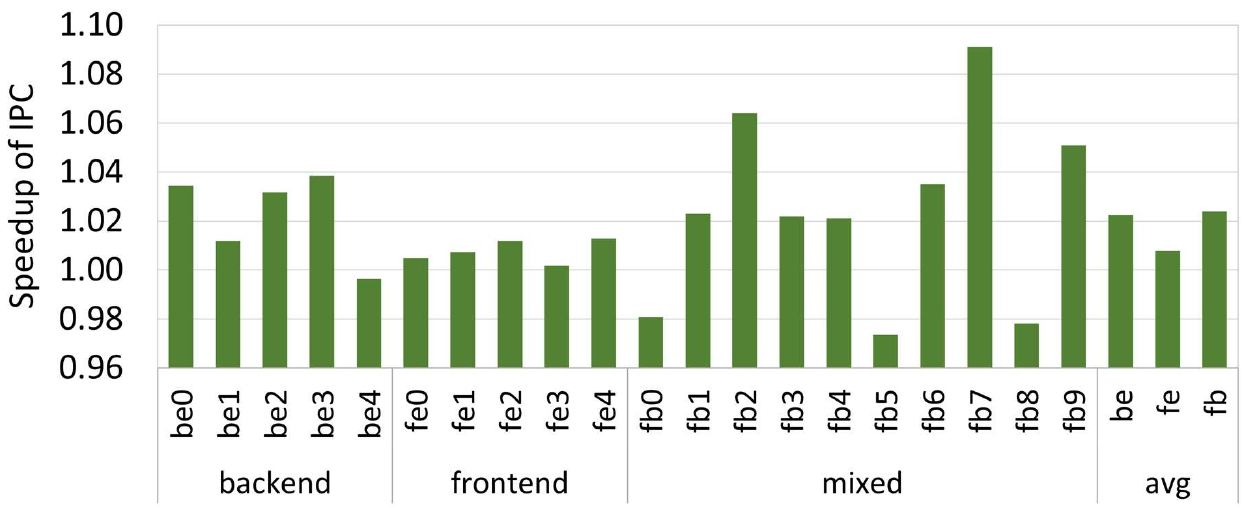}%
\caption{
Speedup of IPC (geomean) over Linux.}
\label{fig:geomean} 
\vspace{-0.5cm}
\end{figure}

Finally, Figure \ref{fig:geomean} presents the IPC speedup results of SYNPA over Linux. 
We compute the IPC of the workloads using the geometric mean of the IPCs of the application in the workload. As observed, IPC speedup values are lower than those of TT
, but again, the mixed workloads are the ones achieving the best speedups, which are, on average, about 2.2\%. As expected, frontend-intensive workloads present similar results to Linux,  with an average speedup of around 0.8\%.

\section{Conclusions}
This paper has presented SYNPA, the first thread allocation policy that is able to work constrained to ARM's performance counters. SYNPA implements a three-category linear regression model computed at the dispatch stage. We found that this stage is the only one where overall system performance can be categorized in ARM processors. The categories were defined after a rough search of the deployed performance counters. 
The devised model guides SYNPA to select the best pair of synergistic applications to be run in each processor core.


SYNPA covers a gap in current thread allocation models as no one can work constrained to ARM's performance counters. Compared to Linux, experimental results show that SYNPA improves TT in random workloads
up to 55\% in some workloads and around 36\% on average.
In addition, fairness is improved up to 25\%. 

Finally, we would like to remark that SYNPA could be integrated as a part of the OS scheduler of current ARM processors, as all the performance counters used in this work are in the standard ARM v8.1 PMU. Of course, the regression model should be trained for the workloads to be run on the target system.



\bibliographystyle{IEEEtran}
\bibliography{biblio}

\end{document}